\documentclass[%
aps,
twocolumn,prd,
showpacs,preprintnumbers,nofootinbib,
amsmath,amssymb,
aps,
superscriptaddress]{revtex4-1}

\usepackage{graphicx}
\usepackage{dcolumn}
\usepackage{bm}
\usepackage{hyperref}
\usepackage{color}
\usepackage{float}

\begin{document}

\title{Detectability of gravitational waves from binary black holes:\\Impact of precession and higher modes
}

\affiliation{Center for Relativistic Astrophysics and School of Physics, Georgia Institute of Technology, Atlanta, GA 30332}

\author{Juan Calder\'on~Bustillo$^\text{}$}\noaffiliation
\author{Pablo Laguna$^\text{}$}\noaffiliation
\author{Deirdre Shoemaker $^\text{}$}\noaffiliation


\preprint{LIGO-P1600257}

\newcommand{\ds}[1]{{\color{magenta}[#1]}}
\newcommand{\jc}[1]{{\color{blue}[#1]}}

\begin{abstract}
Gravitational wave templates used in current searches for binary black holes omit the effects of precession of the orbital plane and higher order modes. While this omission seems not to impact the detection of sources having mass ratios and spins similar to those of GW150914, even for total masses $M > 200M_{\odot}$; we show that it can cause large fractional losses of sensitive volume for binaries with mass ratio $q \geq 4$ and  $M>100M_{\odot}$, measured the detector frame. For the highest precessing cases, this is true even when the source is face-on to the detector.
Quantitatively, we show that the aforementioned omission can lead to fractional losses of sensitive volume of $\sim15\%$, reaching $>25\%$ for the worst cases studied. Loss estimates are obtained by evaluating the effectualness of the SEOBNRv2-ROM double spin model, currently used in binary black hole searches, towards gravitational wave signals from precessing binaries computed by means of numerical relativity. We conclude that, for sources with $q \geq 4$, a reliable search for binary black holes heavier than $M>100M_\odot$ needs to consider the effects of higher order modes and precession. The latter seems specially necessary when Advanced LIGO reaches its design sensitivity.
 \end{abstract}

\maketitle


\section{Introduction}
The recent direct detection of the gravitational wave (GW) signals GW150914 \cite{Abbott:2016blz} and GW151226 \cite{Abbott:2016nmj} by the Advanced LIGO detectors \cite{TheLIGOScientific:2014jea} has initiated the era of GW astronomy. GW signals are determined by the parameters of their sources, and the related science will crucially depend on our ability to measure these parameters. According to current GW models, both analytical \cite{Hannam:2013oca,Pan:2013rra,Babak:2016tgq,Abbott:2016izl, Abbott:2016nmj, TheLIGOScientific:2016wfe} and numerical  \cite{Abbott:2016apu}, the detected GW signals were emitted by two coalescing binary black holes (BBHs) with respective total masses $M= (65.3^{+4.1}_{-3.4}, 21.8^{+5.9}_{-1.7})M_{\odot}$, which implied the simultaneous discovery of BBHs \cite{Abbott:2016nmj}. Remarkably, both binaries showed low mass ratios, respectively bounded by $q<(1.53, 3.57)$, and no compelling evidence for a precessing orbital plane.

Searches for GWs from BBHs are based on the matched filtering of the incoming data to theoretical models of the emitted GW signal,  known as templates\cite{MatchedFilter,Babak:2012zx,Usman:2015kfa,TheLIGOScientific:2016qqj}. Thus, successful detection and identification of BBHs depend crucially on the accuracy of these templates. Existing waveform models include those computed within the framework of effective-one-body (EOB) \cite{Buonanno:1998gg,Taracchini:2012ig,Taracchini:2013rva,Purrer:2014fza} and phenomenological \cite{Ajith:2007qp,Ajith:2007kx,Santamaria:2010yb} formalisms. Both need to be calibrated to numerical relativity (NR) simulations \cite{SXS,Mroue:2013xna, Szilagyi:2009qz,Scheel:2008rj, Frauendiener:2011zz, Hannam:2010ec,ChuThesis,Jani:2016wkt,Vaishnav:2007nm, Herrmann:2006ks,Healy:2009zm} to faithfully model the late stages of the coalescence. Despite their success in detecting the GW signals GW150914 and GW151226, current searches only implement versions of these models which lack two physical phenomena: the precession of the orbital plane of the binary and the so called higher order modes (HMs) of the GW emission (see Table I). In the past it has been shown that omission of HMs can cause significant event losses for the case of non-precessing  \cite{Capano:2013raa,Pekowsky:2012sr,Bustillo:2015qty}, and non-spinning \cite{Varma:2014jxa} BBHs with mass ratio $q \geq 4$ and total mass larger than $\sim100M_{\odot}$. A recent study including the effects of precession~\cite{Harry:2016ijz} demonstrated that GW searches for BBH with mass ratio $q<5$ and total mass $M < 100M_{\odot}$ would increase their sensitivity to precessing sources if precessing templates were added to the corresponding template bank.  A  follow-up study of GW150914, however,  showed that precession and HMs are not likely to significantly impact the parameter estimation of sources in its immediate vicinity of the parameter space, at the SNR at which GW150914 was detected \cite{Abbott:2016wiq}.\\

In this paper we study the impact of the joint omission of both HMs and precession effects in current and future GW searches for binary black holes. We note that the LIGO LSC recently reported that no GWs from a BBH with $M>100M_{\odot}$ were found during the O1 Science Run \cite{Abbott:2017iws}. For this reason we mostly focus our study on BBHs with total mass $M\geq 100M_{\odot}$ in the detector reference frame. To this end, we test the ability of the SEOBNRv2-ROM double spin waveform model \cite{Taracchini:2012ig,Taracchini:2013rva,Purrer:2015tud}, used in the LIGO O1 search for BBHs \cite{TheLIGOScientific:2016qqj}, to recover NR simulated GW signals including the effects of HMs and precession.  We perform our study in the context of two versions of the Advanced LIGO detector: one indicative of its
early sensitivity, known as early Advanced LIGO (eaLIGO) \cite{Aasi:2013wya} and its design version known as zero-detuned-high-energy-power (zdLIGO) \cite{AdvLIGOcurves}.

\section{Higher modes and Precession}
The GW signal emitted by a BBH in quasi-circular orbit depends on 15 parameters. Eight parameters are intrinsic to the binary system: the individual masses $m_i$ and dimensionless spins $\vec\chi_i$ of its two BHs. We will denote collectively these parameters as $\Xi=(m_1,m_2,\vec\chi_1,\vec\chi_2)$. Three other parameters are the luminosity distance $d_L$ and the angles $\theta$ and $\varphi$ which determine the location of the detector in a reference frame centred in the centre of mass of the binary. The polar angle $\theta$ is defined such that $\theta=0$ coincides with the total angular momentum $\vec J$ of the binary. We will say that a BBH is face-on or edge-on to the observer when the latter is located at $\theta=0$ or $\pi/2$ respectively. Two other parameters are $\iota$ and $\phi$, the angular location of the source in the sky of the detector. Finally, the polarization angle $\psi_p$ and the coalescence time $t_c$ complete the 15 parameters of the binary. In this framework, the strain $h$ produced by a GW signal 
can be expressed as:
\begin{equation}
\begin{aligned}
&h(\Xi;d_L,\theta,\varphi;\iota,\phi,\psi_p;t-t_c)=\\
&=F_+ (\iota,\phi,\psi_p) h_+(\Xi;d_L,\theta,\varphi;t-t_c) +\\ & F_ \times (\iota,\phi,\psi_p) h_\times (\Xi;d_L,\theta,\varphi;t-t_c).
\end{aligned}
\end{equation}
As usual in GW literature, $(h_+,h_\times)$ denote the two polarizations of the GW. The antenna patterns of the detector \cite{Jaranowski:1998qm,Schutz:2011tw} $F_+$ and $F_\times$ can be decomposed in a global factor $F$ and an effective polarization $\psi$ defined as $F=\sqrt{F_+^2 + F_\times^2}$ and $\tan{\psi}=F_\times / F_+$ \cite{Bustillo:2015ova}.
With these definitions, the GW strain can be re-expressed as simply:
\begin{equation}
\begin{aligned}
h=\frac{F}{d_L}\bigg{(}\cos{\psi}\Re({\cal H}) + \sin{\psi}\Im({\cal H})\bigg{)}
\end{aligned}
\label{gwmodes}
\end{equation}\\
Above, ${\cal H}$ denotes the complex GW strain $h_+ + i h_\times$. This can be decomposed as a sum of modes $h_{\ell,m}$ weighted by spin -2 spherical harmonics $Y_{\ell,m}$ as:
\begin{equation}
\begin{aligned}
&{\cal H}=h_+ + i h_\times = \\&
{\cal H} =\sum_{\ell\geq 2}\sum_{m=-\ell}^{m=\ell}Y^{-2}_{\ell,m}(\theta,\varphi)h_{\ell,m}(\Xi;t-t_c).
\end{aligned}
\label{gwmodes}
\end{equation}\\
where $h_{\ell,m}(\Xi;t)=A_{\ell,m}(\Xi;t)e^{-i\phi_{\ell,m}(\Xi;t)}$, $A_{\ell,m}$ and $\phi_{\ell,m}$ being real.\\ 

\begin{figure}[h]
\centering
\includegraphics[width=1.\columnwidth]{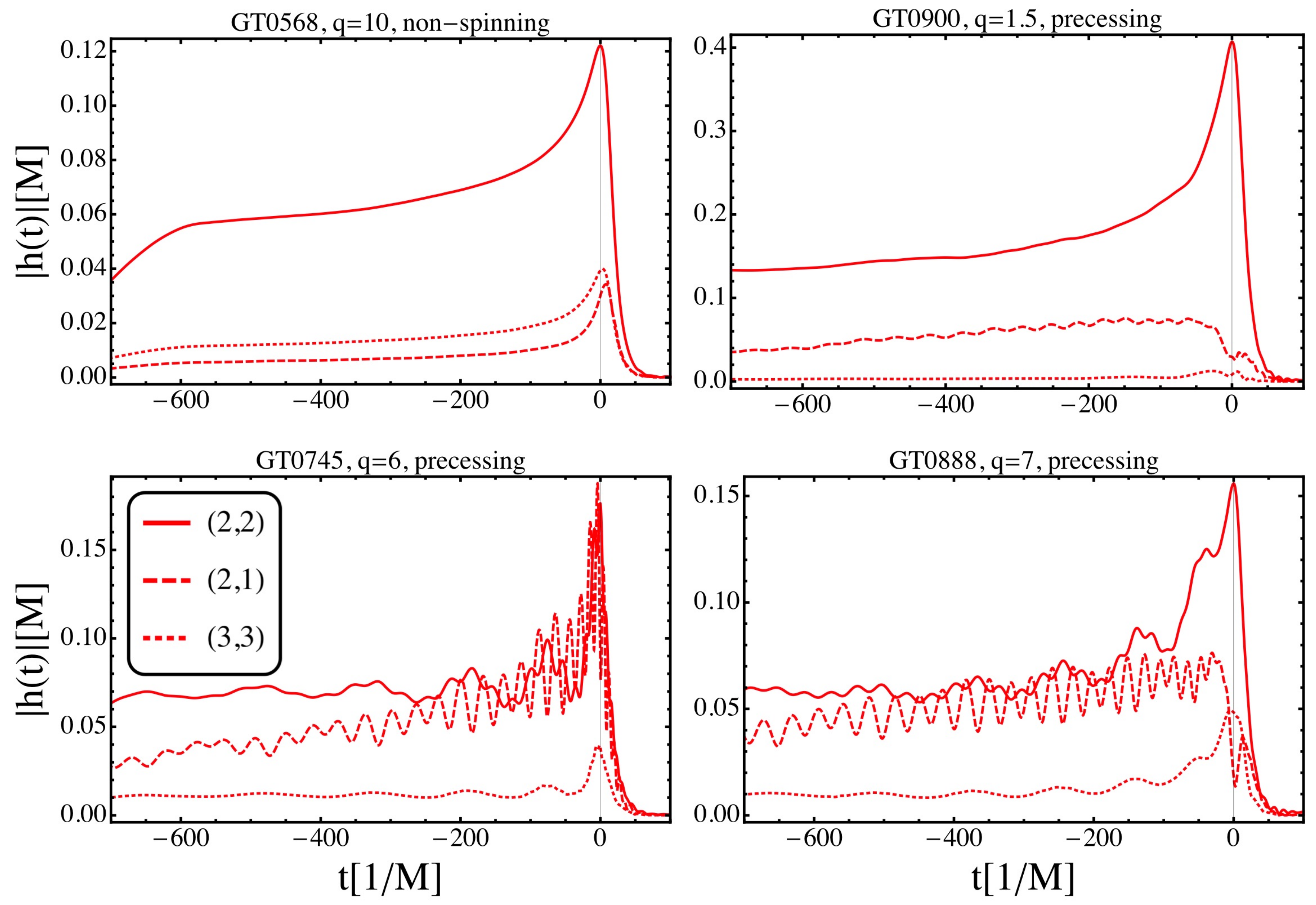}
\caption[Effect of $q$ on higher modes]{We show the amplitude of the dominant $(2,2)$ mode and the $(2,1)$ and $(3,3)$ higher modes for a selection of simulations. First, note how the amplitude  of the $(3,3)$ mode relative to that of the $(2,2)$ mode grows with the mass ratio. Second, note how precession triggers, via mode-mixing, both the amplitude of the $(2,1)$ mode and oscillations in all the modes. These are clearly larger for the two, bottom, large mass ratio cases.}
\label{ex:fig:fws}
\end{figure}

To a very good approximation, the phase $\phi_{\ell,m}(t)$ of each mode is related to the orbital phase of the binary by  
\begin{equation}
\phi_{\ell,m}(t) = m \phi_{orb}(t).
\end{equation}
For the case of non-precessing sources, the GW strain is dominated by the $(\ell,m)=(2,\pm 2)$ modes, reason why current GW searches for BBHs omit the contribution of the HMs \cite{TheLIGOScientific:2016qqj, Usman:2015kfa}. As shown in Fig. \ref{ex:fig:fws}, the HMs only contribute significantly to the GW signal during the last few cycles and merger of nearly edge-on binaries, i.e., $\theta \rightarrow \pi/2$. Their effect is also enhanced as the total mass $M=m_1+m_2$ and the mass ratio $q=m_1/m_2$ grow (where $q \geq1$) \cite{Berti:2007fi,Capano:2013raa,Pekowsky:2012sr,Bustillo:2015qty,Bustillo:2015ova}.\\

\begin{figure*}[ht]
\centering
\includegraphics[width=1.00\columnwidth]{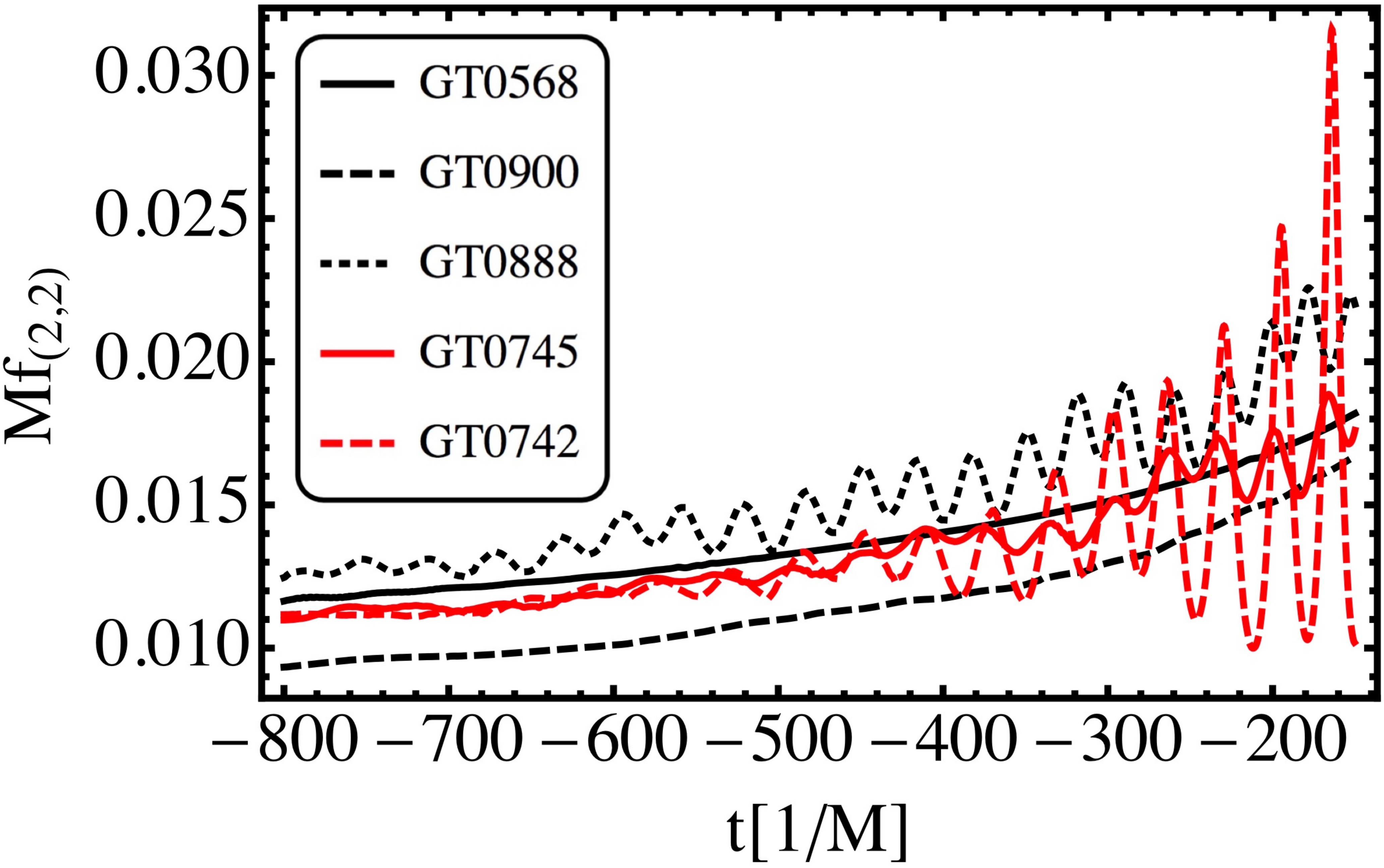}
\includegraphics[width=.97\columnwidth]{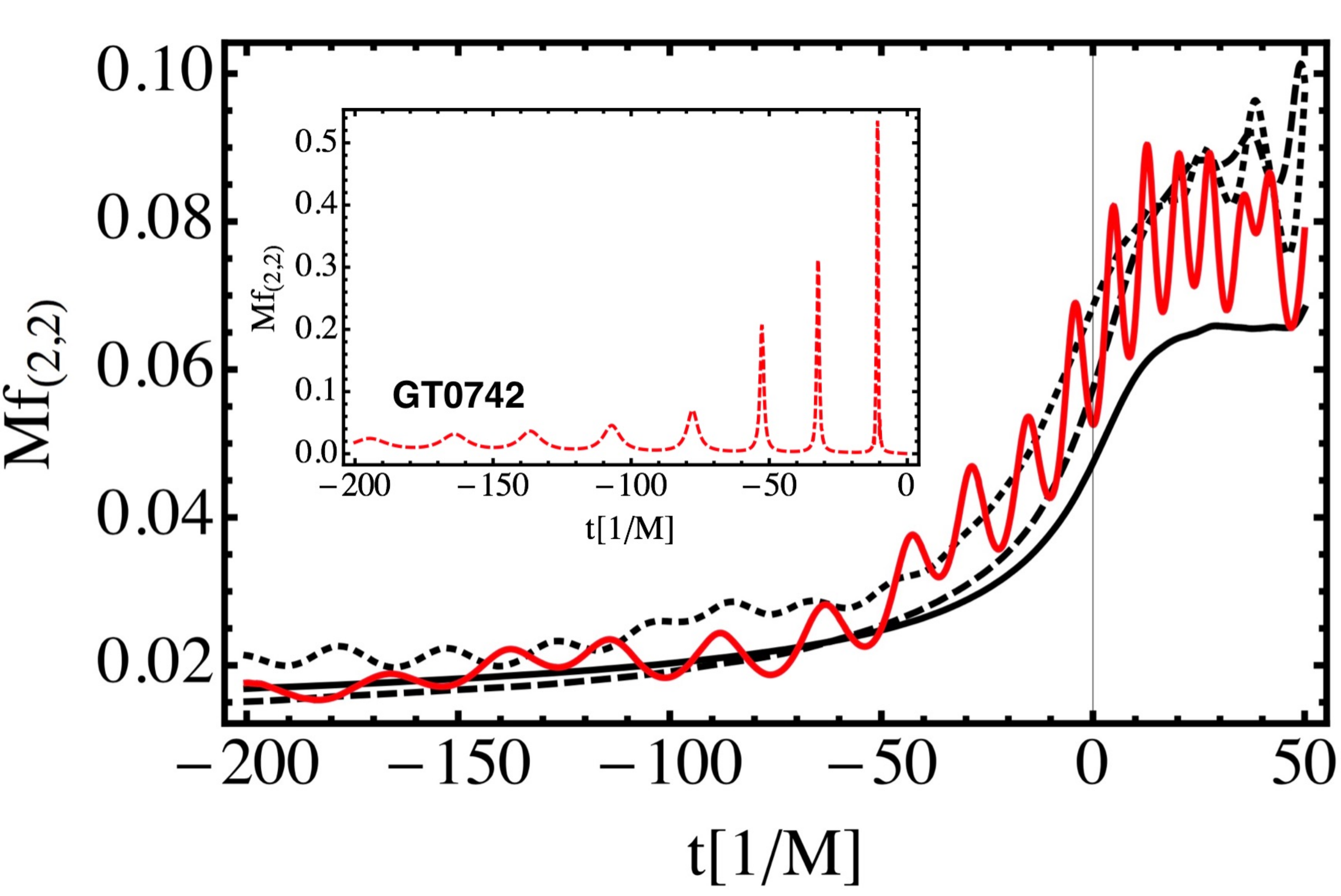}
\caption[Effect of $q$ on higher modes]{Frequency, in geometrical units, of the $(2,2)$ modes of the four simulations shown in Fig.\ref{ex:fig:fws} and the GT0742 case. Note that the the frequency of the 4 precessing cases show oscillations that are not present in the non-spinning GT0568 case, which are specially strong for the cases GT0745 and GT0742. The left panel shows the lower frequency end, in which oscillations are already clear for the latter two cases. The right plot zooms in the high frequency end, where oscillations become more violent. This is particularly noticeable for of GT0742.}
\label{ex:fig:freqs}
\end{figure*}

The total angular momentum $\vec{J}$ of the binary can be expressed as a sum of the orbital angular momentum $\vec{L}$ and the two BH spins as:
\begin{equation*}
\begin{aligned}
\vec{J}= \vec{L} + \vec\chi_1 m_1^2+\vec\chi_2 m_2^2 \equiv \vec{L}+ \vec{S},
\end{aligned}
\label{LJ}
\end{equation*}
Non-zero spin components within the orbital plane of the binary make $\vec{L}$ precess around $\vec{J}$, leading to a precessing orbital plane \cite{Apostolatos:1994mx,Kidder:1995zr}. It is common to respectively group the in-plane and parallel components of the
BH spins into the two parameters $\chi_p$ and $\chi_{eff}$, defined by\footnote{Here, $\vec\chi_i=\vec\chi_i^{||}+\vec\chi_i^{\perp}$, with $\vec\chi_i^{||}=\vec{L}\cdot \vec{\chi_i}$ denoting the parallel component of $\vec\chi_i$.} \footnote{Here, $A=2+1.5 q$.} \cite{Ajith:2009bn, Schmidt:2014iyl}:
\begin{equation*}
\begin{aligned}
\chi_p=\max \left(\chi_{1}^\perp,\frac{A}{q^2} \chi_{2}^\perp\right) \qquad  \qquad  \chi_{eff}=\frac{\chi_1^{||} m_1 + \chi_2^{||} m_2 }{m_1 + m_2}\,.
\end{aligned}
\label{gwmodes}
\end{equation*}
A non-zero $\chi_p$ characterises the presence of a precessing orbital plane, which induces modulations of the frequency and amplitude of the $h_{\ell,m}$ modes, when these are computed in the non-precessing frame of reference \cite{Schmidt:2012rh,Schmidt:2010it,Ossokine:2013zga,Apostolatos:1994mx,Kidder:1995zr}. The latter also leads to an effective mode-mixing: the modes computed in the non-precessing frame are a combination of those computed in the co-precessing one. This leads to precessing waveforms showing particularly strong $(2,\pm 1)$ modes, mainly due to the contribution of the $(2,2)$ mode of the co-precessing frame. Fig.\ref{ex:fig:fws} shows the amplitude of $(2,1)$, $(2,2)$ and $(3,3)$ modes as a function of time, in geometric units. Note how a larger $q$ induces larger $(3,3)$ modes, this being specially large for the case of the non-spinning $q=10$ binary. Note also how, unlike in the case of the non-spinning binary, precessing modes show oscillations in the amplitude and how a large mass ratio triggers both these oscillations and the amplitude of the $(2,1)$ mode. For the particular case of GT0745, we can see how the $(2,1)$ mode even dominates the $(2,2)$ close to the peak of the emission. Fig.\ref{ex:fig:freqs} shows the frequency of the $(2,2)$ modes of these binaries. Note how the precessing cases show frequency oscillations, particularly strong for GT0745 and GT0742, as noted in \cite{Schmidt:2010it}, while the non-spinning case GT0568 shows the typical monotonically growing frequency of non-precessing binaries.\\

Table \ref{models}. provides an overview of the most up to date waveform models. We stress that while some of these include the effects of either precession \cite{Hannam:2013oca,Pan:2013rra,Babak:2016tgq} or HMs \cite{Pan:2011gk}, they are not currently used for detection purposes \cite{Abbott:2016izl}.
Together, both HMs and precession make the signal $h$ differ from the one predicted by non-precessing models that consider only the dominant  $h_{2,\pm 2}$ modes, potentially damaging the ability of current searches to detect GW signals from BBHs.

\begin{table}
\centering
    \begin{tabular}{|l|l|l|l|l|l|l|}
  \hline
    Model & HMs & $\text{Spin}$ & $\text{Precession}$ & Det. & PE & Refs.\\ \hline
   
 SEOBNRv2  & No &    Yes  & No & Yes     & Yes  & \cite{Taracchini:2012ig,Taracchini:2013rva, Buonanno2013} \\ \hline
  SEOBNRv4  & No &    Yes  & No & Yes     & Yes  & \cite{Bohe:2016gbl} \\  \hline
   SEOBNRv3 & $\ell \leq 2$ &   Yes  & Yes & No    & Yes  & \cite{Pan:2013rra,Babak:2016tgq} \\ \hline
         IMRPhenomD  & No &    Yes& No & Yes     & Yes &  \cite{Khan:2015jqa,Husa:2015iqa}  \\ \hline
    IMRPhenomP  & $\ell \leq 2$ &    Yes  & Yes & No     & Yes   & \cite{Hannam:2013oca}\\ \hline
        EOBNRv2HM  & Yes &    No & No & No     & Yes  & \cite{Pan:2011gk} \\ \hline
            \end{tabular}
    \caption{Summary of most up to date waveform models, the effects they include and the purposes they are available for. Here, Det. denotes detection and PE denotes parameter estimation. We want to stress that both precessing models include the $l \leq 2$ modes.}
    \label{models}
\end{table}

\section{Method}
\subsection{Analysis}
The core of searches for BBH is the matched filtering of the detector output signal $s$ to gravitational waveform template banks $\cal B$ composed of templates $h$. The output signal $s$ is in general a superposition of a GW signal $g$ and detector background noise $n$, which we will consider Gaussian and with zero mean. Under this assumption, the output of the matched filter,  known as the signal-to-noise ratio (SNR), is directly related to the probability that $h$ is buried in $s$. The overlap ${\cal O}(h|g)\in [0,1]$ of a template $h$ to a given waveform $g$ is proportional to the SNR that the template can extract from $g$. Similarly, one can define the fitting factor or effectualness of a template bank $\cal B$ to $g$ as the overlap maximised over the templates $h_i$ of $\cal B$ \cite{Apostolatos:1995pj}. 
The overlap (and the fitting factor) is maximal and equal to $1$ when $h=g$ and decreases when as $h$ differs from the waveform $g$, reducing the recovered SNR. For this reason, it is crucial that template banks implement templates that are accurate representations of the incoming GW signals\\ 
 
In this study we aim to assess the ability of current searches to recover GW signals emitted by precessing binaries via studying the effectiveness of the waveform models that its template banks implement \cite{Bustillo:2015qty,Varma:2014jxa}. To this end we compute the fitting factor of our simulated NR precessing signals to the waveform model implemented in the template bank of the LIGO O1 BBH search \cite{TheLIGOScientific:2016qqj}. The latter is a reduced-order-model (ROM) of the SEOBNRv2 waveform model, which neglects the effects of precession and HMs. Unlike template waveforms used in similar previous studies\cite{Capano:2013raa,Bustillo:2015qty,Varma:2014jxa} this model includes unequal spin dynamics, which adds an extra degree of freedom that might improve signal recovery\footnote{We want to emphasise that, as in previous similar studies, we are not running any  detection pipeline used by the LIGO LSC. Instead, we are only testing the ability of the waveform model implemented in these pipelines for the O1 BBH search to recover our target signals}.\\

For a given simulation in Table II. and a given total mass $M$, we compute the fitting factor ${\cal F}_i$ and the optimal SNR $\rho_i^{opt}$ at a fiducial effective distance $d_L/F$. We do this for a collection of $2500$ orientations and effective polarizations $ (\theta_i,\varphi_i,\psi_i)\equiv \Lambda_i$ of the binary, and compute the orientation-averaged fitting factor $\cal F$ as in \cite{Bustillo:2015qty,Varma:2014jxa}. With this, we can compute the optimal and suboptimal volumes $V$ in which a source can be detected at a given fiducial SNR, and the corresponding fractional volume loss as \footnote{We note that in \cite{Bustillo:2015ova} it is defined $\Delta V [\%] = 100\times {\cal F} ^3$. However, volume losses are then plotted and quoted as ($100-\Delta V)$.}:
\begin{equation}
\begin{aligned}
\Delta V [\%] = &100 \times \frac{V_{opt}-V_{subopt}}{V_{opt}}\\ & = 100\times (1-{\cal F} ^3),
\end{aligned}
\end{equation} 
where
\begin{equation}
{\cal F}= \bigg{(}\frac{\sum_j {\cal F}_{i}^3 \rho_{i}^3}{\sum_j \rho_{i}^3}\bigg{)}^{1/3}.
\end{equation}
Above, the $i$ index runs over the different orientations and effective polarizations $\Lambda_i$. For a detailed description of the above method, we invite the reader to look at \cite{Varma:2014jxa, Bustillo:2015ova}.\\

As a final remark, we recall that template banks used in GW searches for BBHs, aim for maximum losses of $10\%$ of signals \cite{TheLIGOScientific:2016qqj,Brown:2012nn}. This implies that the minimum overlap between neighbour  templates of the template bank is $ \sim 0.965$ \cite{Ajith:2012mn}. Instead, we compute the fitting factor of our NR target signals to a continuous waveform family, effectively using a ``perfect" template bank with a minimum overlap of $1$ between neighbour templates. For this reason, the losses we estimate are lower bounds of the ones in which real template banks would incur.

\subsection{Target signals}
We consider as a model of the true GW signal NR waveforms computed by Maya-ETK code \cite{Vaishnav:2007nm, Herrmann:2006ks,Healy:2009zm}, recently presented in the GeorgiaTech NR catalogue \cite{Jani:2016wkt}, that include the HMs and precession.  From this catalog of NR waveforms,  we choose three classes of sources:
\begin{enumerate}
\item Four sources with $q \leq 2$ (as GW150914 has been found to have) and various spin parameters $\chi_{eff}$ and $\chi_p$. We expect HMs to be negligible for these cases due to the low mass ratio.
\item Nine sources with larger $q$ and varying spin parameters. In this case, we expect both HMs and precession to impact the signal.
\item We choose two non-spinning cases with $q=1$ and $q=10$. For these, only the HMs impact the signal, which will allow us to compare our results to previous studies.
\end{enumerate}

These NR simulations are summarised in Table \ref{injections}, 
we scale them to total masses $M\in[70,220]M_{\odot}$. As in similar studies \cite{Varma:2014jxa,Bustillo:2015qty}, we use detector frame parameters. We note that cosmological effects are expected to impact the signals emitted by sources within the Advanced LIGO sensitive range \cite{Martynov:2016fzi}, leading to the detector frame signal being redshifted with respect to the one emitted by the source. This implies that the total mass of a source located at a redshift $z$ will differ from that measured in the detector frame by:
\begin{equation}
M_{detector}=M_{source} (1+z).
\end{equation}
For instance, for the two GW detections GW150914 and GW151226, it was found $z \simeq 0.1$ \cite{Abbott:2016blz, Abbott:2016nmj}.\\

 As mentioned before, we use frequency cutoffs of $f_0=30$Hz for eaLIGO and $f_0=24$Hz for zdLIGO. We note that although the latter is predicted to be sensitive down to $f_0=10$Hz, our NR waveforms do not reach such low frequencies for masses $<M\simeq210M_{\odot}$ for most cases.  In order to avoid the need of creating hybrid NR+post-Newtonian waveforms \cite{Hannam:2007ik,Santamaria:2010yb,Ohme:2011zm,MacDonald:2011ne,MacDonald:2012mp,Bustillo:2015ova}, we start at $f_0=24$Hz. Finally, we note that we start the signals in time domain at the time the $(2,2)$ mode enters the detector band, i.e., the moment $t_0$ in which its frequency satisfies $f_{2,2}(t_0)=\frac{1}{2\pi}\frac{d\phi_{2,2}}{dt}(t_0)=f_0$. In doing this, and since from Eq.(3) $f_{\ell,m} (t) \simeq \frac{m}{2} f_{2,2}(t)$, we are neglecting small contributions from HMs with $m > 2$, which enter the detector band before the $(2,2)$ mode does.\\

\section{Results}

As shown in Fig.1, precessing sources show $(2,2)$ modes whose amplitudes and frequencies differ from the monotonously growing ones of non-precessing targets. When precessing effects are strong enough, we expect them to lead to low $\cal{F}$ even when the source is face-on and the HMs are negligible \cite{Bustillo:2015qty, Pekowsky:2012sr}. As the inclination $\theta$ of the source increases, HMs contribute more to the target signal, leading in principle to further discrepancies between our NR signals and SEOBNRv2. However, as we will see, HMs do not always lead to lower $\cal{F}$. In particular, we observe this is not the case when $\cal{F}$ is already low for face-on orientation.\\



\begin{figure}[h]
\centering
\includegraphics[width=1.02\columnwidth]{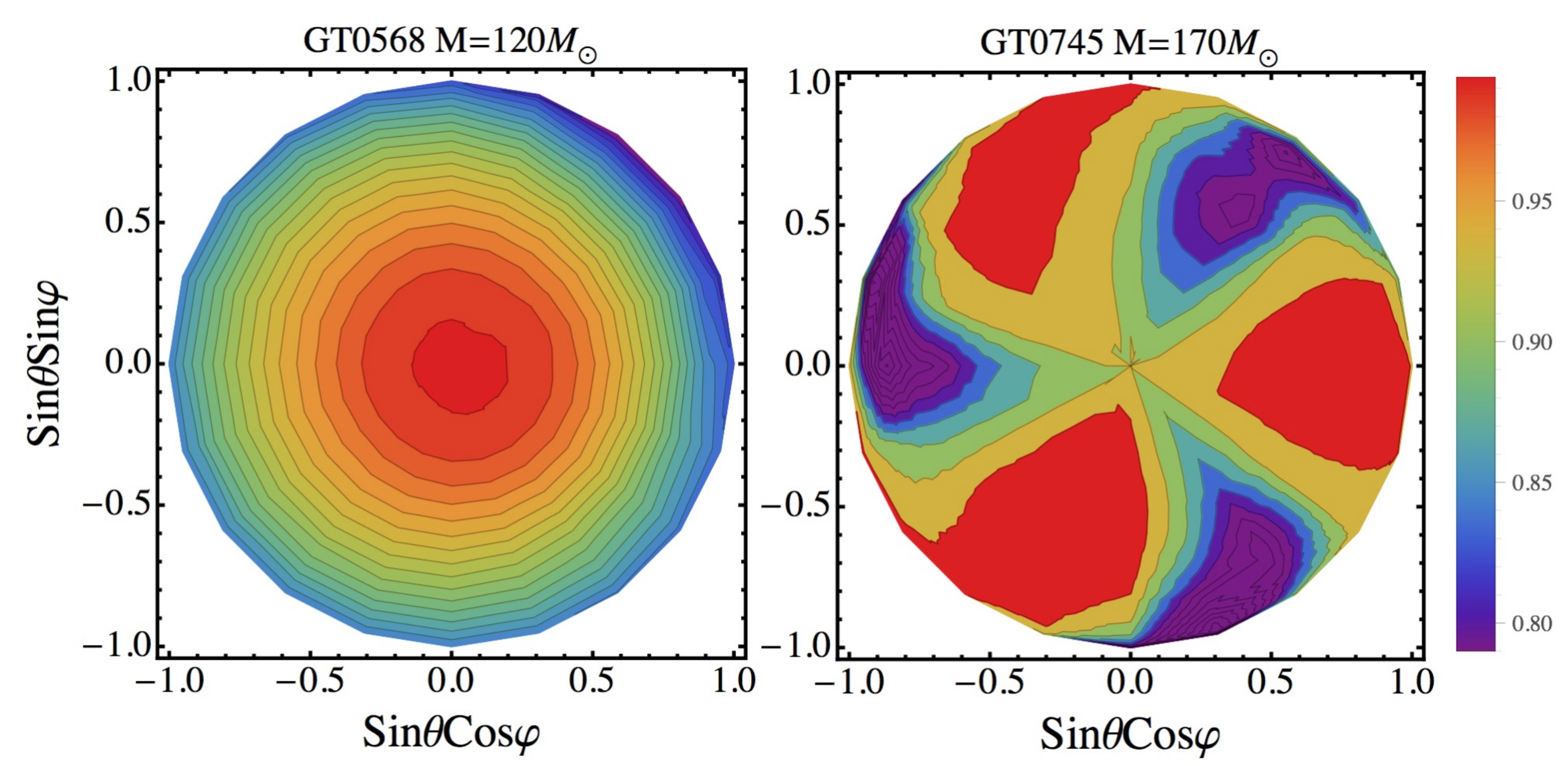}
\caption[Effect of $q$ on higher modes]{Fitting factor, for the case of zdLIGO, of SEOBNRv2 towards two selected NR simulations as a function of $(\theta,\varphi)$, projected onto the $\theta=0$ plane: the non-spinning $q=10$ GT0568, where only HMs impact  the signal, and the precessing $q=6$ GT0745, where precession also contributes. The centre of the plot corresponds to face-on orientation while its contour corresponds to edge-on. Results are shown for orientations with $\theta \leq \pi/2.$}
\label{ex:fig:ff}
\end{figure}

º
\begin{table*}
\centering
    \begin{tabular}{|l|l|l|l|l|l|l|l|}
  \hline
    SIM ID & q & $\vec{\chi_1}$ & $\vec{\chi_2}$ & $\chi_{eff}$ & $\chi_p$ & $ M\omega_0 $ & $M_{min}$\\ \hline
   
    GT0448  & 1 &    (0,0,0)  & (0,0,0) & 0        &  0 &   0.048     &  $70 M_{\odot}$ \\ \hline
 
    GT0455   & 1.2 & (0.42,0,0.42)  &  (0,0,0.6)    & 0.50 & 0.36   & 0.054       &  $150 M_{\odot}$   \\ \hline
    GT0900   & 1.5 & (0.4,0,0)  &  (0.4,0,0)    & 0&  0.4  & 0.048       &  $70 M_{\odot}$   \\ \hline
GT0810   & 2 & (-0.49,-0.05,-0.49)  &  (0,0,0)    & -0.33 & 0.25   & 0.125       &  $180 M_{\odot}$   \\\hline
       GT0825   & 2 & (-0.21,0.44,-0.49)  &  (0,0,0)    & -0.33 & 0.48   & 0.125       &  $180 M_{\odot}$   \\\hline
     GT0430   & 4 & (0,0,0.6)  &  (-0.6,0,0)    & 0.48 & 0.35   & 0.062       &  $100 M_{\odot}$   \\\hline
     GT0438   & 4 & (0,0,-0.6)  &  (-0.6,0,0)    & -0.48 & 0.35   & 0.069       &  $110 M_{\odot}$   \\\hline
     GT0557   & 4 & (-0.6,0,0)  &  (-0.6,0,0)    & 0& 0.6   & 0.063      &  $120   M_{\odot} $ \\ \hline
     GT0560   & 4 & (0.6,0,0)  &  (0,0,0)    & 0& 0.6   & 0.063      &  $120   M_{\odot} $ \\ \hline
    GT0889  & 6 &    (0.42,0,0.42)  & (-0.42,0,-0.42) & 0.30           &  0.42 &    0.061  & $100 M_{\odot} $  \\ \hline
            GT0745  & 6 &    (-0.05,0.54,0.26)  & (0,0,0) & 0.22           &  0.54 &    0.057  & $120 M_{\odot} $  \\ \hline
    GT0888  & 7 &    (0.42,0,0.42)  & (-0.42,0,-0.42) &       0.31     & 0.42  &    0.060  & $130 M_{\odot} $  \\ \hline
     GT0886  & 7 &    (0.42,0,0.42)  & (-0.42,0,-0.42) & 0.32           &  0.42 &    0.061  & $110 M_{\odot} $  \\ \hline
     GT0742  & 8 &    (-0.03,0.60,-0.01)  & (0,0,0) & 0.01           &  0.60 &    0.061  & $110 M_{\odot} $  \\ \hline
    GT0568  & 10 &    (0,0,0)  & (0,0,0) & 0           & 0  &    0.069  & $110 M_{\odot} $  \\ \hline
     \end{tabular}
   
    \caption{Summary of Georgia Tech NR simulations used as target signals. $M\omega_0$ denotes the initial frequency of the $(2,2)$ mode of the simulation, in geometric units. The source parameters are defined at this frequency. $M_{min}$ denotes the minimum mass to which each simulation has been scaled for zdLIGO. The $(\ell,m)$ modes were read using the Mathematica script provided in \cite{mathscript}.}
    \label{injections}
\end{table*}
\begin{figure*}[ht]
\centering
\includegraphics[width=0.99\columnwidth]{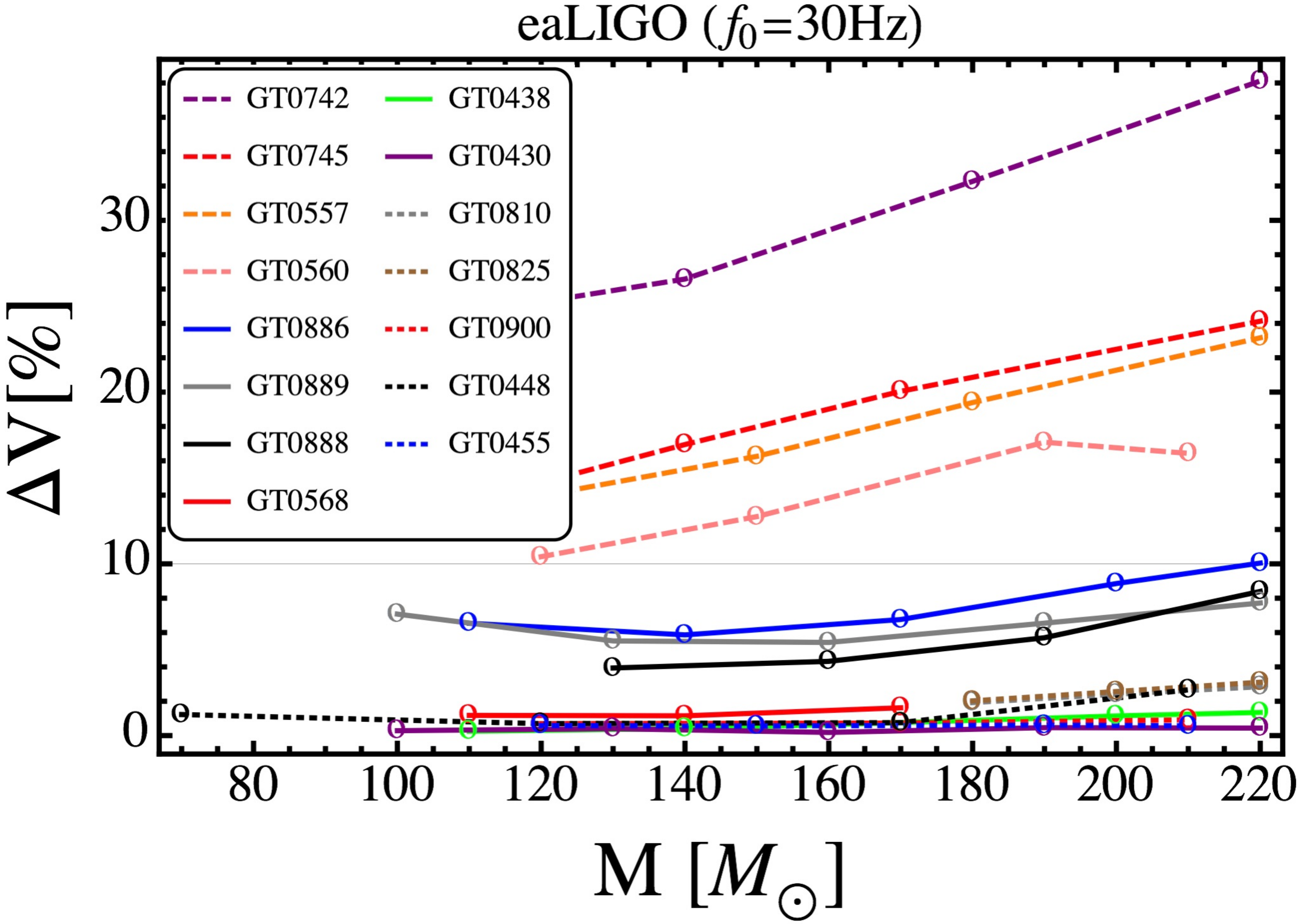}
\includegraphics[width=0.99\columnwidth]{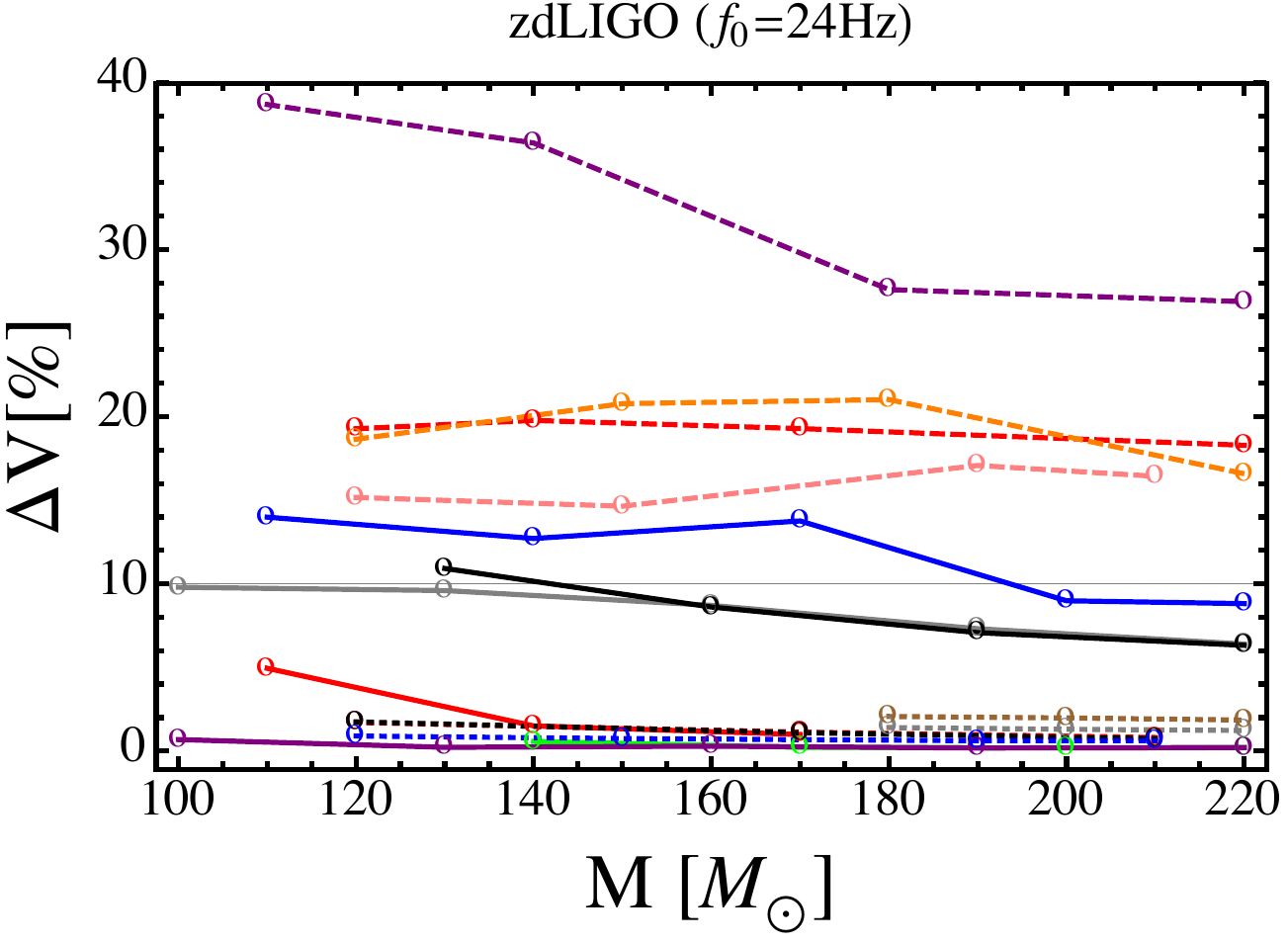}
\includegraphics[width=1.0\columnwidth]{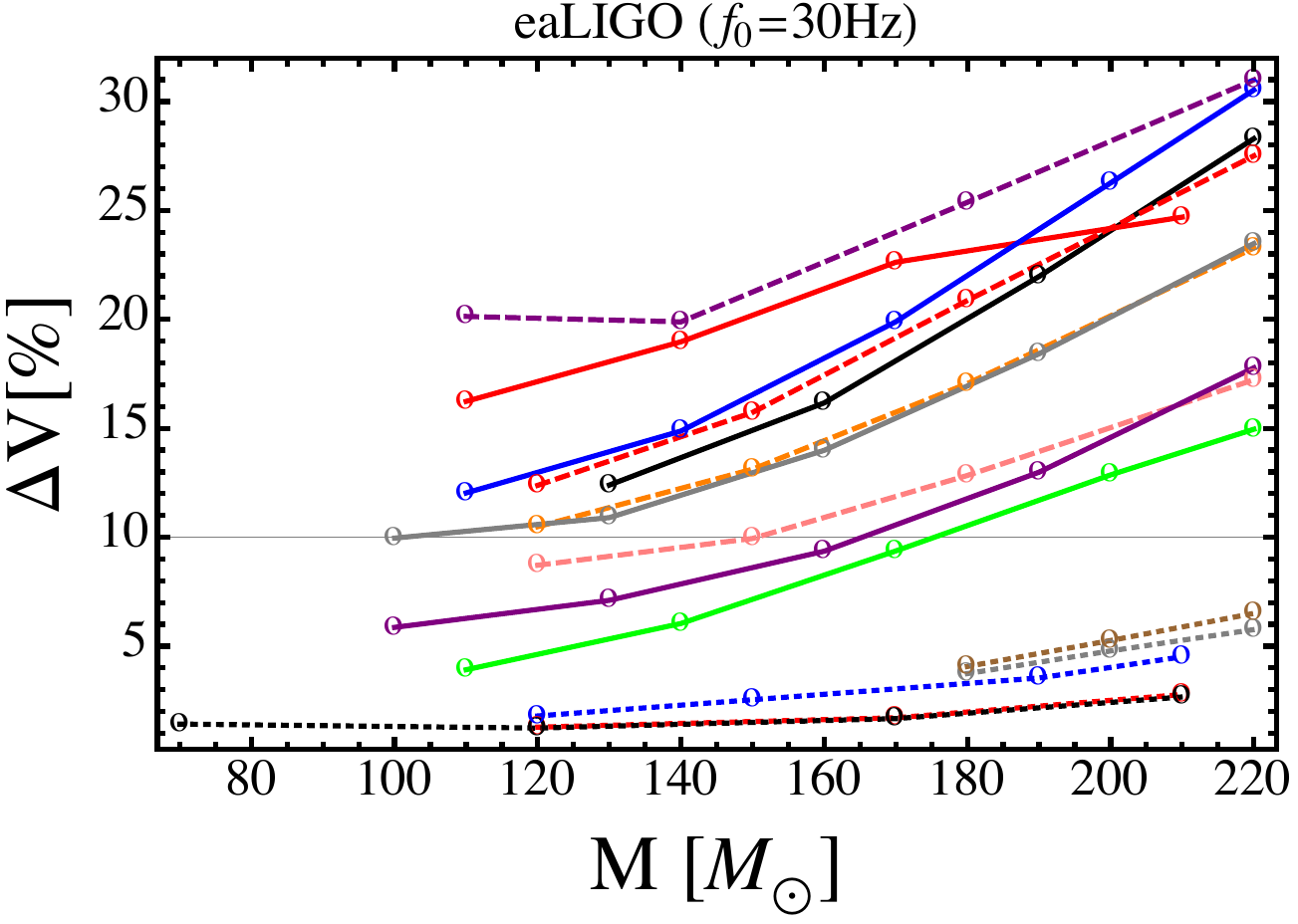}
\includegraphics[width=1.0\columnwidth]{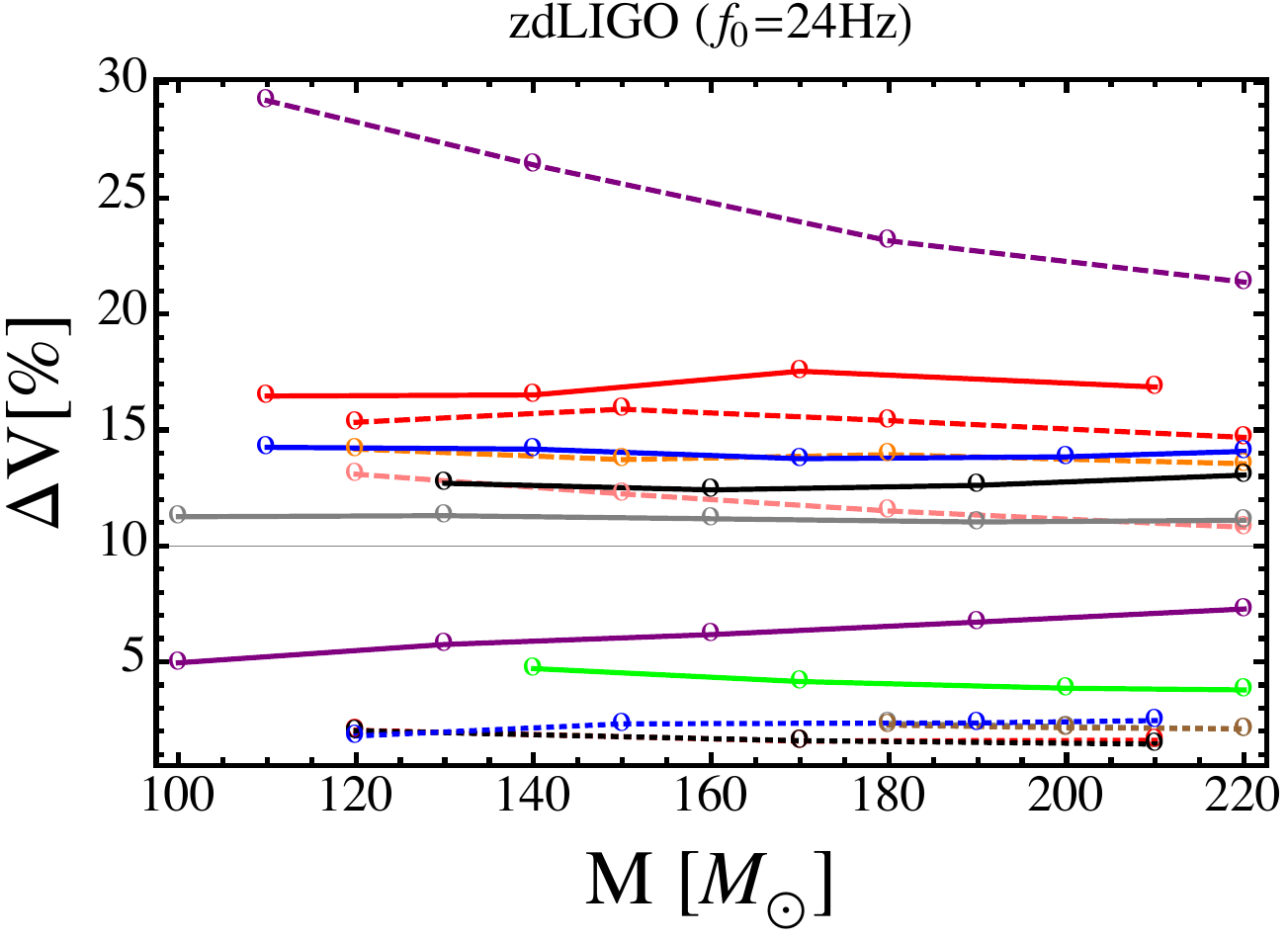}
\caption[Effect of $q$ on higher modes]{Bottom: fractional volume loss for the studied target signals averaged over orientations. The horizontal line denotes the $10\%$ threshold for admissible losses, following the standard criterion of BBH searches. Top: same for the case of the sources being face-on, so that only the $(2,2)$ mode contributes, except for marginal contributions from the $(3,2)$ mode.}
\label{ex:fig:avgff}
\end{figure*}

Fig.~\ref{ex:fig:ff}, shows the ${\cal F}_i$ obtained for two selected simulations as a function of the orientation $(\theta,\varphi)$, averaged over $\psi$: a non-precessing one with strong HMs and a strongly precessing one with strong HMs. Concretely, the left panel corresponds to the $q=10$ non-spinning case, for which HMs dominate the signal, while the right one corresponds to the precessing $q=6$ case, for which  precession also impacts the signal.
When face-on, the non-spinning signal is perfectly matched by SEOBNRv2 due to the absence of HMs. In particular we obtain  ${\cal F}_0= 0.997$, while precession leads to a lower ${\cal F}_0= 0.912$ in the right panel case. Looking at the modes of GT0745 in Figs.\ref{ex:fig:fws} and \ref{ex:fig:freqs}, it is not surprising that the non-precessing SEOBNRv2 model cannot recover such a highly oscillating $(2,2)$ mode. In the first case, ${\cal F}_i$ tends toward lower values as we approach edge-on orientations. In the second, a pattern forms mainly due to the interaction of the $(2,2)$ mode with a strong $(2,1)$ mode, which leads to several local maximums for particular values of $(\theta,\varphi)$.
Note that in both cases ${\cal {F}}_i$ can get below $0.85$ for edge-on systems.\\

\subsection{Face-on results: impact of precession}
The top panels of Fig.~\ref{ex:fig:avgff} show the fractional volume loss for the case in which the sources are face-on to the detector. We recall that in this case, only the $(2,2)$ mode has significant contributions to the GW signal, together with marginal contributions from the $(3,2)$ mode. Thus, any disagreement between our NR targets and SEOBNRv2 is dominated by the impact precession in the $(2,2)$ mode (see Fig.2 and 3). Given this, we can roughly distinguish two cases here, those that show strong precession effects and hence larger losses, and those that show low losses. 
\begin{enumerate}
\item Note that all the cases showing losses above $\sim (5\%,10\%)$ for (eaLIGO,zdLIGO) correspond to precessing sources with $q \geq 4$ in which the heavier BH has non-zero in plane spin components. Among them, note that the binaries having $\chi_p > 0.5$ (dashed lines) clearly show larger losses than those for which $\chi_p < 0.5$ (solid lines). For the former, losses are larger for zdLIGO due to the increased number of precessing cycles in band. Similar to the case of GT0745 mentioned above, note how the extremely oscillating frequency evolution of GT0742 in Fig.2 leads to large disagreements with SEOBNRv2, producing extremely large losses that can surpass $30\%$. 
\item On the other hand, all the other sources showing losses below $5\%$ are either non-precessing, have low mass ratio, or have a heavier BH with no in-plane spin components (dotted lines). As before, all of these sources have $\chi_p < 0.5$, which indicates that this parameter is working well at evaluating the impact of precession for the cases we study.\\
\end{enumerate}

Notice also that while losses increase with the total mass for the case of eaLIGO, they remain quite constant for the case of zdLIGO or even decrease. This is due to the different shapes of the corresponding sensitivity curves: while eaLIGO shows a clear sweet-spot at $\sim 150$Hz, the zdLIGO curve is almost equally sensitive at all frequencies spanned by the waveforms  \footnote{See, for example, the top panels of Fig.4 in \cite{Bustillo:2015qty}.}. The frequency of the GW emission scales as $1/M$, so that increasing the total mass of the binary has the effect of reducing the number of cycles in band and shift the frequency of the GW merger emission, close to which precession effects are the largest. For eaLIGO, this implies that losses will be maximum when the merger frequency is close to that of its sweet-spot (same will apply later for the case of HM's). On the other hand, for zdLIGO, lowering the total mass would increase the number of cycles in band while keeping the merger cycles in a fairly equally sensitive zone of the detector band. This causes losses to remain quite constant or even increase for some cases due to the increased number of precessing cycles in band.\\

\subsection{Orientation averaged results: impact of higher modes}

The bottom panels of Fig.~\ref{ex:fig:avgff} show averaged fractional volume losses for each of the studied sources for eaLIGO (right panel) and zdLIGO (left panel). Two cases are present here now. 

\begin{enumerate}

\item Large averaged losses: First, the sources with $q \geq 4$ have strong HM contributions, which leads to averaged fitting factors differ from the face-on ones. Let us divide them into two sub-cases: 

\begin{itemize}

\item $\chi_p < 0.5$ (solid lines): Due to the low impact of precession, these sources show losses below $\sim 10\%$ when face-on. However, HMs make these increase for other orientations of the source. In particular note how losses dramatically rise in the large mass end for the case of eaLIGO, due to the HMs entering the sweet-spot of the detector as the $(2,2)$ starts leaves the detector sensitive band. Note how the maximum losses are now close to $30\%$ while they barely reached $10\%$ when face-on.

For the case of zdLIGO, losses also increase when HMs are added to these cases, except for GT0886, which showed losses of $\sim15\%$ when face-on.

\item $\chi_p > 0.5$ (dashed lines): Again, the contribution from HMs makes losses differ from the face-on ones. However, their qualitative impact is not as clear as in the previous case. In fact, in some cases, addition of HMs tend to reduce losses with respect to the face-on case. This is for example the case of GT0745, which is consistent with Fig.3 showing maximum fitting factors for orientations other than face-on. In all cases, the losses are still clearly above $10\%$ for both detectors, reaching $25\%$  for some cases. Consistently with \cite{Bustillo:2015ova}, losses are in general larger for the case of eaLIGO, due to the enhanced effect of HMs for the shorter signals it is sensitive to, due to the larger $f_0$. The exception to this is the GT0742 case, for which its strong precession effects (see Fig. 2) are enhanced for zdLIGO, due to its lower $f_0$ increasing the number of cycles in band. 

\end{itemize}

\item Low averaged losses: lets focus on the non-spinning $q=1$ case and the precessing $q\leq 2$ cases, which can be considered heavier versions of GW150914. These sources are always well recovered by the SEOBNRv2 model and always show averaged losses way below $10\%$ for both the face-on and orientation averaged cases. This implies that, in average, none of precession or HMs should have an impact when searching for this sort of systems, the latter being expected due to the low mass ratio of the sources. Consistently, it was recently found that these two effects do not impact the estimation of the parameters of sources  similar to GW150914, for most of the orientations of the source \cite{Abbott:2016wiq}.

\end{enumerate}

Note that for all cases, the averaged losses clearly grow as a function of the total mass for eaLIGO, due to the $\ell > 3$ modes entering its sweet-spot. Note that from Eq.(4), the loss for a given $M_{source}$ at a given redshift $z$, would be, in terms of the losses we show $\Delta V(M_{source},z)=\Delta V (M_{detector} (1+z))$. Since for eaLIGO losses clearly increase with the total mass, these would be larger if we had used source frame masses.\\

Unlike for eaLIGO, in the case of zdLIGO the losses remain fairly more constant along the mass range due the absence of a sweet-spot. These two different behaviours are consistent with those shown in  \cite{Bustillo:2015qty}, Fig.6. Also, in contrast with that study, focused on aligned-spin sources, we find here that most precessing cases show losses in the low mass end which are larger for zdLIGO, due to the increased number of cycles in band impacted by precession. This suggests that precession will impact more the searches in the low mass of our parameter space when Advanced LIGO reaches its design sensitivity.\\

Finally we check that the results we obtain for non-spinning binaries are consistent with previous studies. On one hand, as expected, the non-spinning $q=1$ source is perfectly recovered by SEOBNRv2. On the other, the $q=10$ case shows losses just below $20\%$ for zdLIGO and between $16\%$ and $24\%$ for eaLIGO, consistent with \cite{Bustillo:2015qty,Varma:2014jxa}

\section{On estimating the sensitivity of a search via fitting factors}

To a first order, the sensitivity of a search to a given source can be estimated, as has been done in this and previous studies \cite{Bustillo:2015qty,Varma:2014jxa}, by means of the fitting factor of the corresponding template bank to the GW emitted by the source. 
However, we want to stress that as pointed in \cite{Capano:2013raa,Harry:2016ijz}, a search including HMs or precession would require an increased number of templates. This would raise the False Alarm Rate of the search and require GW triggers to have a larger SNR to be as significant as in the search omitting the aforementioned effects. 
As an example, the study conducted in \cite{Capano:2013raa} shows that triggers having a SNR of $\rho=8$ in a non-spinning search would need to get to $\rho=8.3$ for being equally significant when higher order modes are included in the search, which implies that the source needs to be located at $0.963$ times the original distance. A similar situation is reported for the case of a search for BBH with $(q,M)<(5,100M_{\odot})$, when precession effects are added to an aligned-spin search \cite{Harry:2016ijz}. This effect has not yet been quantified for the case of including HMs and precession to an aligned-spin search, and its estimation is above the scope of this work.\\

We want to also note again that while our SEOBNRv2 bank is effectively infinitely dense (minimum match =1), the aligned spin banks used in real GW searches \cite{TheLIGOScientific:2016qqj} consider a minimum match of $\sim 0.965$. This means that the fitting factors we find are upper bounds of the ones that would be obtained if using a search template bank. Hence, in this sense, the losses we obtain are lower bounds of the real ones.\\

Finally, since we do not run any actual search algorithm, we have also omitted the impact of signal based vetoes that real GW searches implement \cite{Allen:2004gu,Babak:2012zx,HannaThesis}. The goal of these vetoes is to discriminate between triggers produced by real GW signals and spurious triggers of terrestrial origin known as glitches \cite{Blackburn:2008ah,Slutsky:2010ff,Canton:2013joa}. This is done by lowering the significance of signal triggers whose morphology differ from that of the template \cite{Babak:2012zx,Usman:2015kfa}, or in other words, have a low fitting factor with the template that picked them. This would further reduce the sensitivity of current aligned-spin searches to our NR signals, which include the impact of precession and HMs.

\section{Conclusions, limitations and future work}
Current gravitational wave searches for binary black holes implement waveform models that omit the effects of both higher order modes and the precession of the orbital plane of the binary. We have estimated the impact of this omission via computing the fitting factor of such a waveform model towards simulated NR signals including both higher modes and precession effects. We find that the model we consider, SEOBNRv2, shows a good fitting factor toward our NR signals when $q \leq 2$. This suggests that after averaging over orientations, none of precession or higher modes should have an impact in the detection of systems similar to GW150914, even when the total mass surpasses $200M_{\odot}$. This is consistent with the fact that the two binary black holes detected by LIGO up to date, GW150914 and GW151226, exhibited low mass ratios $q < (1.53,3.57)$. In fact, a recent follow-up study for GW150914 \cite{Abbott:2016wiq} showed that neither precession nor higher modes were likely to affect the parameter estimation similar systems, consistently with our results.\\ 

However, we find that for the systems heavier than $100M_\odot$ and $q \geq 4$ considered in this study, searches can be severely compromised by the aforementioned omission. Among these sources, we find that for those having a precessing spin parameter $\chi_p > 0.5$ precession dramatically impacts the resulting GW signal, and hence the fitting factors, even when the sources are face-on. Quantitatively, we observe that omission of precession and HMs effects can lead to orientation-averaged losses of $\Delta V > 25\%$ for the highest masses studied for the case of eaLIGO. Typical losses for zdLIGO are of the order of $\sim15\%$ for all the studied mass range.  We also find that, unlike in the case of non-precessing sources \cite{Bustillo:2015qty}, for the  $q \geq 4$ precessing sources, zdLIGO shows larger losses than eaLIGO in the low mass end of our parameter space, suggesting that consideration of precession effects will be specially needed once Advanced LIGO reaches its design sensitivity, and the length of the signal in the detector band increases.\\ 

Our study is limited by the length of our NR simulations, which required a frequency cutoff $f_0=24$Hz for the case of zdLIGO, far from its expected $f_0=10$Hz value.  As mentioned before we also neglect the impact of signal based vetoes \cite{Allen:2004gu,Babak:2012zx,HannaThesis}, which would further punish signals having a low match to the non-precessing bank waveforms, leading to larger losses. It is also out of the scope of this work to compute the false alarm rate due to the incremented number of templates needed in a search including precession and higher modes. It is been shown, however, that even with this raise the sensitivity of a search would still increase when adding precessing waveforms \cite{Harry:2016ijz}.\\

Altogether, we hope to further motivate the development of complete inspiral-merger-ringdown waveform models that include the effects of both precession and higher order modes and their implementation in searches for BBH.  We note that an effort toward the implementation of a precessing search is already on its way \cite{Harry:2016ijz}.\\

\section{Acknowledgements}
We thank Tito Dal Canton, Sebastian Khan and Karan P. Jani for very useful discussions. We also thank Alex Nielsen, Aycin Aykutalp and Ricardo Martinez-Garcia for comments on the manuscript. JCB, PL and DS gratefully  acknowledge support from the NSF grants 1505824, 1505524, 1333360, XSEDE  TG-PHY120016. JCB thanks the support from the Max Planck-Prince of Asturias Mobility Award, and thanks the hospitality of AEI Hannover, where part of this study was conducted. A Mathematica implementation of the SEOBNRv2 ROM waveform model was kindly provided by Michael P\"{u}rrer. 

\bibliography{HMbib}

\end{document}